\documentclass{emulateapj}

\shorttitle{DSC and the evolution of massive galaxies}
\shortauthors{Monaco et al.}

\begin{document}

\title{Diffuse stellar component in galaxy clusters 
and the evolution of the most massive galaxies at $z\la1$}
\author{Pierluigi Monaco$^{1,2}$}
\author{Giuseppe Murante$^{3,1}$}
\author{Stefano Borgani$^{1,2,4}$}
\author{Fabio Fontanot$^5$}
\affil{$^1$Dipartimento di Astronomia, Universit\`a di Trieste, via Tiepolo 11, 34131 Trieste -- Italy}
\affil{$^2$INAF-Osservatorio Astronomico di Trieste, via Tiepolo 11, 34131 Trieste -- Italy}
\affil{$^3$INAF-Osservatorio Astronomico di Torino, Strada Osservatorio 20, 10025 Pino Torinese (TO) -- Italy}
\affil{$^4$INFN-National Institute for Nuclear Physics, Trieste -- Italy}
\affil{$^5$Max Planck Institute for Astronomy, K\"onigstuhl 17, D-69117 Heidelberg, Germany}

\begin{abstract}
The high end of the stellar mass function of galaxies is observed to
have little evolution since $z\sim1$.  This represents a stringent
constraint for merger--based models, aimed at explaining the evolution
of the most massive galaxies in the concordance $\Lambda$CDM
cosmology.  In this Letter we show that it is possible to remove the
tension between the above observations and model predictions by
allowing a fraction of stars to be scattered to the Diffuse Stellar
Component (DSC) of galaxy clusters at each galaxy merger, as recently
suggested by the analysis of N-body hydrodynamical simulations. To
this purpose, we use the {\sc morgana} model of galaxy formation in a
minimal version, in which gas cooling and star formation are switched
off after $z=1$. In this way, any predicted evolution of the galaxy
stellar mass function is purely driven by mergers.  We show that, even
in this extreme case, the predicted degree of evolution of the high
end of the stellar mass function is larger than that suggested by
data.  Assuming instead that a significant fraction, $\sim 30$ per
cent, of stars are scattered in the DSC at each merger event,
leads to a significant suppression of the predicted evolution, in
better agreement with observational constraints, while providing a
total amount of DSC in clusters which is consistent with recent
observational determinations.

\end{abstract}

\section{Introduction}

The $\Lambda$CDM model provides the standard framework to study the
formation of cosmic structures, with only residual uncertainties on
the values of cosmological parameters.  However, while consensus on
the agreement between model and data is reached for observables that
probe the large-scale structure of the Universe (e.g., Springel, Frenk
\& White 2006), the situation becomes far less clear when the
formation and evolution of galaxies are addressed.  In this case the
underlying astrophysical processes at play are so complex and poorly
understood that it is very difficult to disentangle the
cosmologically-driven building of structure from the effects of such
processes.

At variance with the behaviour of Dark Matter (DM) halos, the building
of galaxies shows a ``downsizing'' or ``anti-hierarchical'' behaviour:
at low redshift the specific star-formation rate is higher for smaller
galaxies, while more massive galaxies show higher specific
star-formation rates at higher redshift (see, e.g., Cowie et al. 1996;
Bundy et al. 2006). Besides, stars in more massive objects appear to
have formed on average earlier than those in less massive ones (see,
e.g., Treu et al. 2005; Thomas et al. 2005). While for the bulk of
galaxies this behaviour can be explained as due to the effect of
stellar or AGN feedback (see, e.g., Croton et al. 2006; Bower et
al. 2006), the nearly passive evolution of the most massive galaxies
highlights a possible paradox of present models of galaxy formation.
More specifically, galaxies with stellar masses $\sim 10^{12}$
M$_\odot$ show a remarkably constant number density out to redshift
$z\sim1$ (see, e.g., Fontana et al. 2004; Drory et al. 2005; Yamada et
al. 2005; Zucca et al. 2005; Caputi et al. 2006; Bundy et al. 2006;
Fontana et al. 2006; Wake et al. 2006; Cimatti, Daddi \& Renzini 2006;
Brown et al. 2006; but see also Bell et al. 2004; Faber et
al. 2005).  These exceptionally massive galaxies are the giant
ellipticals which typically represent the dominant galaxies of rich
galaxy groups and clusters.  Furthermore, galaxy clusters are the most
massive DM halos at low redshift and are predicted and observed to be
still undergoing a phase of significant merger events.  The massive
ellipticals that reside at the centres of two merging clusters
are predicted to merge after one dynamical friction time, which is of
order of 1 Gyr. This leads to two important consequences, namely an
evolution of the stellar mass function, which is constrained by data,
and mergers between big ellipticals. These are not associated to
starbursts, due to the lack of cold gas supply in the merging galaxies
(``dry mergers''), and are rather difficult to observe (van Dokkum
2005, Masjedi et al. 2006; Bell et al. 2006).

On the other hand, galaxy clusters are pervaded by a Diffuse Stellar
Component (DSC), which only in part can be associated with the
extended halo of a dominant cD galaxy. These stars are usually not
accounted for in the census of the stellar mass budged in
clusters. Their number and mass can be estimated by observing
intra-cluster planetary nebulae (Arnaboldi et al. 2002, 2004;
Feldmeier et al. 2003, 2004a) intracluster novae and supernovae
(Gal-Yam et al. 2003; Neill et al 2005), AGB stars (Durrell et
al. 2002) using surface photometry of single clusters (Gonzalez et
al. 2000; Feldmeier et al. 2002; Feldmeier et al. 2004b; Krick et
al. 2006) or by measuring the diffuse light in coadded images of many
galaxy clusters (Zibetti et al. 2005).  These observations give
fractions of total luminosity contributed by the DSC ranging from 10
to 40 per cent in massive clusters. The relatively poorer Virgo and
Fornax clusters have observed fractions of about 10 per cent
(Feldmeier et al. 2003; Durrell et al. 2003; Neill et al 2005; Mihos
et al. 2005), thus suggesting an increasing DSC fraction with cluster
richness (see also Lin \& Mohr 2004).  The origin of the DSC in galaxy
clusters has been studied with the aid of N-body simulations
(Napolitano et al. 2003; Murante et al. 2004; Willman et al.2004;
Sommer-Larsen et al.2005; Rudick et al. 2006; Stanghellini,
Gonz\'alez-Garc\'ia \& Manchado 2006), reaching the general
conclusion that a DSC is naturally expected to arise from the
hierarchical assembly of clusters.  In particular, Murante et
al. (2006) showed that 60 to 90 per cent of the DSC is generated at $z
<1$, and only a minor part of it is due to tidal stripping, the rest
being contributed by relaxation processes during galaxy mergers.

Clearly, the possibility that a significant amount of stars are
diffused into the DSC during the low--redshift ``dry assembly'' of the
most massive ellipticals has important consequences of the
evolution of the high--mass end of the galaxy stellar mass function.
Massive galaxies at the centre of clusters contain a significant
fraction of the total stellar mass of the cluster, ranging from
$10-30$ per cent for poor clusters ($M_{\rm h}\sim10^{14}$ M$_\odot$)
to $5-10$ per cent for rich ones ($M_{\rm h}\sim10^{15}$ M$_\odot$;
see, e.g., Lin \& Mohr 2004). If at each merger these galaxies lost a
fair fraction of their stars to the DSC component, and if this
mechanism were responsible for the build-up of most of the DSC, then
this process would limit the mass growth of the central galaxy
by mergers since $z\sim1$.

In this Letter we show, using the results of N-body simulations and
the {\sc morgana} galaxy formation model (Monaco, Fontanot \& Taffoni
2006), that the evolution of massive galaxies driven by mergers is
severely constrained by observations, and that this tension is removed
if a significant fraction of stars is lost to the DSC at each merger.
Once this effect is taken into account we predict a much slower
evolution of the high end of the stellar mass function at $z\la 1$,
while producing an amount of DSC at $z\sim0$ which is consistent with
current observational limits.  In this paper we use a cosmology with
$\Omega_0=0.3$, $\Omega_\Lambda=0.7$, $\Omega_b=0.04$, $H_0=70$ km
s$^{-1}$ Mpc$^{-1}$, $\sigma_8=0.9$; none of the results depends
sensitively on any of these parameters.

\section{Building of the Diffuse Stellar Component}
\begin{figure}
\epsscale{1.}
\plotone{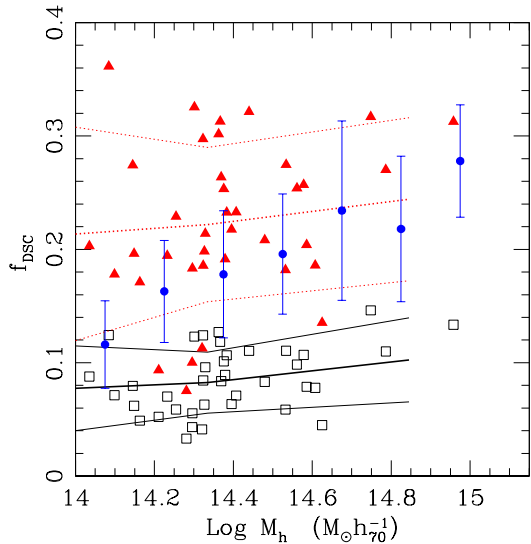}
\caption{Fraction of stars in Diffuse Stellar Component for
model DM halos with $M_{\rm h}>10^{14}$ M$_\odot$.  The black open
squares refer to the expectation of {\sc morgana} with only tidal
stripping ($f_{\rm scatter}=0$), the red triangles to the case $f_{\rm
scatter}=0.3$.  The (continuous and dotted) lines give the average
(thick lines) and $\pm1\sigma$ (thin lines) location of the points.
For comparison we show the results from simulations by Murante
et al. (2006) as circles with errorbars, which represent the
r.m.s. scatter within different mass intervals.
\label{dsc}}
\end{figure}

Murante et al. (2004; 2006) analysed hydrodynamical simulations of
galaxy clusters, performed with the {\small GADGET-2} code (Springel
2005), which include the processes of star formation and supernova
feedback.  They found that the DSC represents a significant fraction
of the stellar population in clusters, approximately ranging from 10
to 40 per cent, with an increasing trend with cluster mass (see the
blue points of Fig.~\ref{dsc}), thus in keeping with observational
results.  Murante et al. (2006) also shows that the bulk of the DSC is
not due to tidal stripping of non-central galaxies, which accounts for
no more than 5-10 per cent of the total stellar component, but to
relaxation processes taking place during the dry mergers leading to
the build-up of the central dominant galaxy. As a result, up to $\sim
30$ per cent of the stellar mass of the merging galaxies becomes
unbound to the resulting central galaxy. In terms of the mass of each
merging satellite, this translates in $10-50$ per cent of its mass
which is scattered to the DSC, depending on the mass ratio of the
merging galaxies.

In this Letter we resort to the novel {\sc morgana} model of galaxy
formation to quantify the effect of including the generation of a DSC
at each merger on the evolution of the stellar mass function.  This
code has been shown to be able to reproduce the build-up of the
massive galaxies (Fontana et al. 2006) and the population of AGNs
(Fontanot et al. 2006).  For the purpose of the present analysis, {\sc
morgana} has been modified by switching off gas cooling and star
formation at $z<1$. In this way, we minimize the evolution of the
stellar mass function, which is then driven only by mergers.
Furthermore, we implement the generation of the DSC as follows: (i)
tidal stripping of stars is applied to satellite galaxies\footnote{At
the time of first periastron of the satellite orbit in the host DM
halo, all the stars that lie beyond the tidal radius (according to the
unperturbed profile of the galaxy) are moved to the DSC}; (ii)
when the satellite merges with the central galaxy a fraction $f_{\rm
scatter}$ of its stars are scattered to the DSC.  Prescription (ii) is
at variance with Monaco et al. (2006), where scattering is allowed
only in major mergers.  Such a recipe, inspired by the results of
Murante et al. (2006), is deliberately simplified and we use it here
to provide a qualitative picture of the effect of including the
production of the DSC into our model.

In Fig.~\ref{dsc} we compare the fraction of DSC, $f_{DSC}$, as a
function of cluster mass, found in the simulations analysed by Murante
et al. (2004) and predicted by {\sc morgana} for both $f_{\rm
scatter}=0$ and 0.3. {\sc morgana} predictions have been computed for
37 clusters, with mass $M_{\rm H}>10^{14}$ M$_\odot$, identified in a
150 Mpc box where the DM clustering is sampled with 512$^3$
particles. This comparison shows that using a fixed value of $f_{\rm
scatter}$ produces a milder dependence of $f_{DSC}$ on the cluster
mass, thus confirming that our approach to introduce the effect of the
DSC generation is oversimplified. Still, predictions from the
semi--analytical model and from the hydrodynamical simulations share
several common features. For instance, tidal stripping is confirmed to
bring only $\sim 10$ per cent of the total stellar mass to the DSC,
with $f_{\rm scatter}\sim 0.3$ required to better account for
simulation results. Quite interestingly, we also verified that
$\sim70$\% of the DSC is generated at $z\la1$ by both {\sc morgana}
and simulations. Based on these results, we conclude that the {\sc
morgana} model can be used to test the effect of the DSC generation on
the evolution of the high end of the galaxy stellar mass function.

\section{Results and Discussion}
As mentioned in the Introduction, the population of massive galaxies,
with $M_\star\sim 10^{11}$ M$_{\odot}$, show a modest but significant
degree of evolution since $z\sim1$.  Using the GOODS-MUSIC sample,
Fontana et al. (2006) found this evolution to amount to a factor of
2.5 in mass density, a degree of evolution which has been shown to be
consistent with the predictions of {\sc morgana}.  On the other hand,
very massive galaxies with $M_\star\sim10^{12}$ M$_\odot$, show a much
lower degree of evolution. We use here as a convenient quantification
of this evolution the logarithmic increase of $M_{-4.5}$, the stellar
mass at which the stellar mass function reaches the level $\Phi({\rm
Log} M)=10^{-4.5}$ Mpc$^{-3}$, from $z=1$ to 0. A detailed discussion
on how to measure this quantity from data is beyond the scope of this
paper. Using data from Yamada et al. (2005), Drory et al. (2005),
Bundy et al. (2006), Cimatti et al. (2006), Fontana et al. (2006) and
Brown et al. (2006) we infer that the evolution of
$M_{-4.5}$ {between $z=1$ and 0} cannot be larger than 0.2 dex.  This
modest evolution clearly requires that massive galaxies must have had
a small net gain in stellar mass during the last 7 Gyr.

\begin{figure}
\epsscale{1.}
\plotone{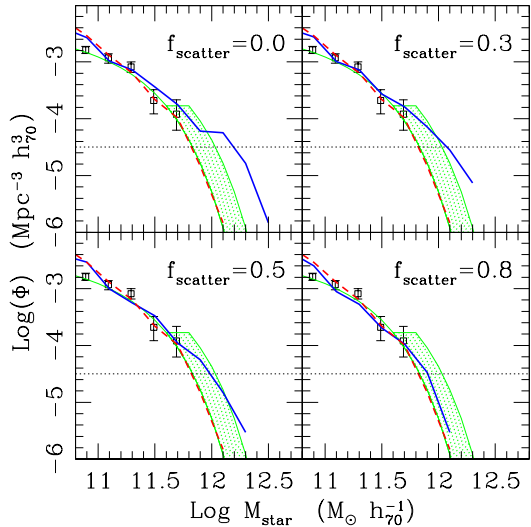}
\caption{Evolution of the stellar mass function from $z=1$ to $z=0$.
In all panels, observational data points are from GOODS-MUSIC (Fontana
et al. 2006) and refer to the stellar mass function in the redshift
range $0.8-1.3$; the green continuous line gives the best fit proposed
by the same authors at $z=1$.  The shaded region highlights the
allowed evolution of the high end of the stellar mass function by 0.2
dex.  The red dashed line gives the model results at $z=1$, computed
assuming $f_{\rm scatter}=0$ and fine-tuned to reproduce very
accurately the analytic fit at the same redshift, while the blue
continuous line gives the prediction at $z=0$, computed switching off
all astrophysical processes (cooling, star formation and feedback) and
setting $f_{\rm scatter}$ to the value specified in the panel.  The
thin dotted horizontal line marks the level $10^{-4.5}$ Mpc$^{-3}$
that is used to quantify the evolution of the stellar mass function.
\label{mfbar_01}}
\end{figure}

To test the consistency of this constraint with the expected
evolution of massive galaxies, we use the {\sc morgana} model as
follows.  We follow the evolution of the galaxy population until
$z=1$, assuming the standard choice of parameters used both in Monaco
et al. (2006) and Fontana et al. (2006) with\footnote{This is done in
order to have all models starting from the same configuration at
$z=1$. As in {\sc morgana} $\sim$70\% of the DSC is created at $z<1$,
we correct our $f_{\rm DSC}$ values by multiplying them by $1/0.7$.}
$f_{\rm scatter}=0$.  We then fine-tune AGN feedback\footnote{The fine
tuning is performed by setting the $f_{\rm jet,0}$ parameter to 2 in
place of 1 and assuming the ``forced quenching'' procedure, see Monaco
et al. (2006) for details.} to reproduce almost exactly the analytic
fit of the $z=1$ stellar mass function proposed by Fontana et
al. (2006).  Fig.~\ref{mfbar_01} shows the predicted mass function at
$z=1$ (dashed line), compared to the GOODS-MUSIC estimate in the
redshift range $0.8-1.3$; the shaded region, bound by the analytic fit
of the observed stellar mass function at $z=1$ and the same curve
shifted in mass by 0.2 dex, highlights the allowed range of the high
end at $z=0$.  The model is known to overestimates at $z=1$ the number
density of smaller objects ($M_\star\la10^{11}$ M$_\odot$, Fontana et
al. 2006), and this is noticeable in the figure.  As already
mentioned in Section 2, we then compute the evolution of the galaxy
population at $z<1$ by switching off all the astrophysical processes,
including cooling, star formation, feedback, galactic winds and
superwinds, so that galaxies can grow only by mergers. The solid line
in the upper left panel of Fig.~\ref{mfbar_01} shows the results of
this model for $f_{\rm scatter}=0$: we obtain $\Delta \log
M_{-4.5}\simeq0.3$, i.e. the mass of the most massive galaxies grows
by more than a factor of two, in line with the results by De Lucia et
al. (2006) and De Lucia \& Blaizot (2006), but at variance with
respect to observational results.

\begin{figure}
\epsscale{1.}
\plotone{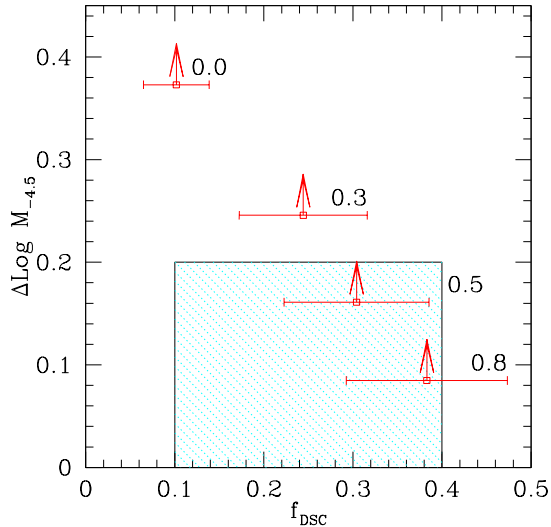}
\caption{A comparison of model and observations in the $f_{\rm
DSC} - \Delta {\rm Log} M_{-4.5}$ parameter space.  It shows the
relation between the production of Diffuse Stellar Component and the
evolution of the stellar mass function at the fixed number density of
$10^{-4.5}$ Mpc$^{-3}$.  The shaded area gives the rough observational
constraints reported in this Letter ($\Delta{\rm Log} M_{-4.5}<0.2$
and $0.1<f_{\rm DSC}<0.4$), the points refer to the model with the
four values of $f_{\rm scatter}$ (reported beside the relative points)
given in Fig.~\ref{mfbar_01}.  We consider these points as lower
limits (see text).
\label{space}}
\end{figure}

This result highlights the presence of a potential paradox in
cosmological models of galaxy formation: even under the assumption
that mergers only drive the evolution of the galaxy population at
$z<1$, model predictions still provide too strong an evolution of the
high end of the stellar mass function. 
This conclusion is robust against possible uncertainties in the
dynamical friction time-scales, which determine the difference between
the timing of DM halo merging and galaxy merging. We verified
that, since these time--scales are much smaller than the Hubble time,
an uncertainty in their estimate does not significantly influence the
final results.

As already discussed in Section 2, the model with $f_{\rm scatter}=0$
also underestimates the fraction of DSC produced in simulations for
the most massive clusters (see Fig.~\ref{dsc}). The other three panels
of Fig.~\ref{mfbar_01} show the evolution of the stellar mass function
for values of $f_{\rm scatter}$=0.3, 0.5 and 0.8. Values between 
0.3 and 0.5 are sufficient to suppress $\Delta \log M_{-4.5}$
to below 0.2 dex, and at the same time reproduce the observed fraction
of DSC.  The rather extreme value of $f_{\rm scatter}=0.8$ instead
tends to over-produce the DSC.

From these results we conclude that the observed modest evolution of
the high-mass tail of the stellar mass function can be reconciled with
model predictions by allowing a significant fraction of the stellar
mass to be scattered away from the galaxies and disperse into the DM
halo. This is shown also in Fig.~\ref{space}, where the results of the
models are reported in the $f_{DSC}$ - $\Delta\log M_{-4.5}$ plane as
lower limits to the values that would be obtained with a full
treatment of baryon physics. The shaded area shows the region
currently allowed by data.  As a word of caution, we remind that a
direct comparison between the theoretical and observational estimates
of the DSC fraction is quite delicate. Theoretical estimates are
affected by numerical effects and by uncertainties in the modeling of
complex baryon physics that give rise to galaxies, while observational
estimates depend upon a number of hypothesis linking the observables
(e.g. number of intra--cluster planetary nebulae, ratio of
fluxes from the DSC and from galaxies) to the volume--averaged
$f_{DSC}$.

Despite all these uncertainties we regards our result as a robust one.
The details of the galaxy formation models are immaterial in this test
as long as the model gives a plausible population of massive galaxies
at $z\sim1$, and describes correctly the merging of galaxies driven by
the hierarchical assembly of DM halos. In our calculation the
evolution to $z=0$ can only be underestimated, since it is performed
by forcing a complete quenching of cooling and star formation.  This
is clearly seen in Fig.~\ref{mfbar_01}, where the population of
galaxies with $M_\star\sim10^{11}$ M$_{\odot}$ is underestimated at
$z=0$.  As a consequence, the evolution predicted by mergers is an
underestimate as well, as it does not include the stars formed since
$z=1$.  In this case the known excess of small galaxies predicted at
$z=1$ (Fontana et al. 2006) gives a modest bias, which is in the
opposite direction with respect of the more important bias obtained by
quenching any evolution of the stellar component. Therefore, it does
not hamper by any means our conclusions.

In conclusion, we have shown that the modest evolution of the
high-mass end of the stellar mass function may highlight a problem for
current models of galaxy formation in the $\Lambda$CDM framework.  On
the other hand, the presence of a significant DSC in galaxy clusters
and the mild evolution of the high end of the galaxy stellar mass
function may both point toward a scenario in which a significant
fraction of the stellar mass of galaxies becomes unbound at each
merging event, thereby suppressing the merger--driven
evolution. Solving this problem requires that a significant fraction,
$>20$\%, of the total stellar budget in rich galaxy clusters must be
in the form of a diffuse component. Deeper searches of intra--cluster
light are necessary to either confirm or dispute this
prediction. Future instruments, like the Large Binocular Camera at LBT
or JWST, will provide a quantum leap in the census of the diffuse
stars in the near future.

\acknowledgments 
We thank Alvio Renzini, Stefano Cristiani, Andrea Cimatti and
Gabriella De Lucia for useful discussions. This work has been
partially supported by the PD-51 INFN grant.

\end{document}